\documentclass{emulateapj}
\usepackage{graphicx}
\usepackage{epstopdf}
\usepackage{natbib}

\slugcomment{Submitted to ApJ}

\newcommand{\xmm}{\textit{XMM-Newton}}

\newcommand{\lsim}{\lesssim}
\newcommand{\NH}{\ensuremath{N_{\rm H}}}
\newcommand{\psrA}{PSR~J0337}
\newcommand{\psrB}{PSR~J0636}
\newcommand{\psrC}{PSR~J0645}

\begin{document}

\title{Ordinary X-rays from Three Extraordinary Millisecond Pulsars:
  XMM-Newton Observations of PSRs J0337+1715, J0636+5129, and J0645+5158}
\author{Ren\'ee Spiewak\altaffilmark{1}, David L.\ Kaplan\altaffilmark{1}, Anne Archibald\altaffilmark{2}, Peter Gentile\altaffilmark{4}, Jason Hessels\altaffilmark{2,3}, Duncan Lorimer\altaffilmark{4}, Ryan Lynch\altaffilmark{5}, Maura McLaughlin\altaffilmark{4}, Scott Ransom\altaffilmark{5}, Ingrid Stairs\altaffilmark{6,7}, and Kevin Stovall\altaffilmark{8}}
\altaffiltext{1}{Department of Physics, University of Wisconsin-Milwaukee, P.O. Box 413, Milwaukee, WI 53201, USA}
\altaffiltext{2}{ASTRON, the Netherlands Institute for Radio Astronomy, Postbus 2, 7990 AA, Dwingeloo, The Netherlands}
\altaffiltext{3}{Anton Pannekoek Institute for Astronomy, University of Amsterdam, Science Park 904, 1098 XH Amsterdam, The Netherlands}
\altaffiltext{4}{Department of Physics, West Virginia University, 210E Hodges Hall, Morgantown, WV 26506, USA}
\altaffiltext{5}{National Radio Astronomy Observatory, 520 Edgemont Road, Charlottesville, VA 22901, USA}
\altaffiltext{6}{Department of Physics and Astronomy, University of British Columbia, 6224 Agricultural Road, Vancouver, BC V6T 1Z1, Canada}
\altaffiltext{7}{McGill Space Institute, 3550 University St., Montreal, Quebec, H3A 2A7 Canada}
\altaffiltext{8}{Department of Physics and Astronomy, University of New Mexico, Albuquerque, NM 87131, USA} 

\begin{abstract}
We present the first X-ray observations of three recently discovered
millisecond pulsars (MSPs) with interesting characteristics: PSR~J0337+1715, PSR~J0636+5129, and PSR~J0645+5158.  PSR~J0337+1715 is a fast-spinning, bright, and so-far unique MSP in a hierarchical triple system with two white dwarf (WD) companions.  PSR~J0636+5129 is a MSP in a very tight 96-min orbit with a  low-mass, 8\,$M_J$ companion.  PSR~J0645+5158 is a nearby, isolated MSP with a very small duty cycle (1-2\%), which has led to its inclusion in high-precision pulsar timing programs.  
Using data from \xmm, we have analyzed X-ray spectroscopy for these three objects, as well as optical/ultraviolet photometry for PSR~J0337+1715.  The X-ray data for each are largely consistent with expectations for most MSPs with regards to the ratios of thermal and non-thermal emission.  
 We discuss the implications of these data on the pulsar population, and prospects for future observations of these pulsars.  
\end{abstract}


\maketitle

\section{Introduction}
The study of millisecond pulsars (MSPs) has led to many discoveries in astronomy and physics.  Because of their extreme nature and precision in radio emission, 
these objects have been used to constrain theories of relativistic gravity 
\citep[e.g.,][]{ksm+06}, and understand pulsar emission \citep[e.g.,][]{fst88}, binary evolution \citep[e.g.,][]{crl+08}, and the equation of state for material at supra-nuclear densities
\citep[e.g.,][]{dpr+10}, and, in the long-term, are being used to constrain and ultimately detect 
gravitational waves with pulsar timing arrays (PTAs; e.g., \citealt{jhvs+06}).  

MSPs are important targets of high-energy (X-ray and $\gamma$-ray) facilities.  \textit{Fermi} has done wonders to revolutionize our understanding of non-thermal emission from pulsar magnetospheres \citep{aaa+09,rrc+11} for energetic (spin-down luminosities $\dot E \gtrsim 10^{34}\,{\rm erg\,s}^{-1}$) pulsars and continues to help identify new energetic MSPs \citep[e.g.,][]{ksr+12}. In contrast, the soft X-ray band (0.2-10\,keV) not only probes energetic pulsars, but also gives vital information about the surface emission of a wider range of MSPs. In soft X-rays, the emission consists of a combination of non-thermal emission from the pulsar magnetosphere and thermal emission from heated polar caps \citep{zavlin07,dkp+12}, with the ratio depending on the pulsar's age, its spin-down luminosity, $\dot E$, and geometric factors \citep[e.g.,][]{pccm02}.  
For the more common MSPs with low spin-down luminosities ($\dot E \lsim 10^{33}\,{\rm erg\,s}^{-1}$), which account for $\approx60$\% of MSPs with $\dot E$ measurements in the ATNF Pulsar Catalogue\footnote{\url{http://www.atnf.csiro.au/people/pulsar/psrcat/}} \citep{mhth05}, studying the dominant thermal emission from individual MSPs has been a powerful probe of neutron star heating and has allowed constraints on the equation-of-state of supra-nuclear matter \citep{bgr08}, and starts to probe surface inhomogeneities and magnetic field geometries.  
\citet{grml+14} compiled data from X-ray observations of 49 MSPs (with periods of $<30$\,ms)\footnote{\url{http://astro.phys.wvu.edu/XrayMSPs/}}, $\approx15\%$ of all pulsars with similar periods, some of which only have distance measurements from dispersion measure models, which have large uncertainties. 
The \textit{Neutron Star Interior Composition Explorer (NICER)}, which is anticipated to launch in 2016, will be able to measure the radii of neutron stars to better than 10\% uncertainty through soft X-ray observations, to experimentally determine the equation-of-state of neutron stars \citep{gao12}.

Discovering interesting new MSPs is the major driver behind 
the Green Bank Telescope Driftscan (GBTDrift; \citealt{blr+13,lbr+13}) and Green Bank North Celestial Cap (GBNCC; \citealt{slr+14}) pulsar surveys. In particular, these 350-MHz surveys 
aim to discover a large number of new MSPs in areas in which MSPs are under-represented, and thereby significantly improve the sensitivity of the International Pulsar Timing Array (IPTA) 
efforts.  GBTDrift was carried out in mid-2007 and covers northern and southern declinations, while the GBNCC survey initially covered the sky north of $\delta=38^\circ$, an area 
which is inaccessible to Parkes and Arecibo. Ongoing observations are now moving to lower declinations while data analysis of the existing data is in process. 
Because of the predominately high latitudes 
and low frequency of these surveys we expect to see proportionally more nearby MSPs than conventional pulsar surveys (i.e., compare \citealt{bjd+06} and \citealt{jbo+09} to \citealt{mlc+01}).  GBTDrift found 31 new pulsars including 7 MSPs, while GBNCC has published
67 pulsars and 9 MSPs\footnote{\url{http://arcc.phys.utb.edu/gbncc}}.   Here we discuss \xmm\ observations of three of the more interesting discoveries from these surveys.  

\object[PSR J0337+1715]{PSR~J0337+1715} (hereafter \psrA) is an MSP in a stellar triple system that was discovered in GBTDrift data; it is the first such system discovered.  
\psrA\ has two white dwarf (WD) companions in hierarchical orbits \citep{rsa+14}, so this system could provide a way to test theories of relativistic gravity 
such as the strong equivalence principle.  The stable nature of this system and the 1.6- and 327-day orbits could also allow us to study 3-body dynamics on a variety of timescales, and understand the formation and evolution of MSP systems \citep{rafikov14,lg14,tvdh14,ss15}.  

\object[PSR J0636+5129]{PSR~J0636+5129} (hereafter \psrB) is an MSP with a very low-mass companion that was discovered in GBNCC data in a 96\,min orbit: one of the tightest MSP binary systems known \citep{slr+14}.  The companion, which has a minimum mass of 7.4\,M$_J$ (for an assumed neutron star mass of 1.4\,M$_\odot$), does not show any signs of current mass loss (cf.\ \citealt{rfs+12}) and appears similar in nature to the ``diamond planet" orbiting PSR J1719--1438 \citep{bbb+11}.  This suggests that \psrB\ will evolve into an isolated MSP following a period of mass-transfer/loss in an ultra-compact X-ray binary \citep{db03,bbb+11,vhnvj12}.  

\object[PSR J0645+5158]{PSR~J0645+5158} (hereafter \psrC) is a nearby, isolated MSP with a duty cycle of only 1--2\% at 820\,MHz, and timing observations at 820\,MHz have provided a timing solution with a residual RMS of 0.51\,$\mu$s, which makes it an excellent addition to the PTAs \citep{slr+14}.  The full width at half-max (FWHM) of the pulse was measured at 820\,MHz to be $86\,\mu$s; according to the ATNF Pulsar Catalogue, only 8 out of 115 MSPs (with recorded FWHM values) have pulse widths $<100\,\mu$s.

In what follows, we scale quantities to the pulsar distances found by \citet{kvkk+14} and \citet{slr+14}: $1300\pm80$\,pc, $210^{+30}_{-20}$\,pc, and $650^{+200}_{-130}$\,pc for PSRs\,J0337, J0636, and J0645, %
respectively.  We note that these are not dispersion measure distances but are based on WD atmosphere models (\psrA) and timing parallax (\psrB, \psrC), and so should be more accurate \citep[e.g.,][]{gmcm08,cbv+09,roberts11}.  However, for \psrC, the low significance of the measurements indicates that it may be slightly biased by sampling effects (see \citealt{vwc+12}).  A more precise distance measure from the VLBA for \psrA\ will be obtained within the year.  

In Section~\ref{ssec:obsx}, we summarize our methods and the spectral models fit to the X-ray data.  In Sections~\ref{ssec:0636} and \ref{ssec:opt}, we discuss our tests for orbital variation in \psrB\ data and look at optical/UV data for \psrA.  In Section~\ref{sec:disc}, we discuss the implications of our findings for the pulsar population.

\begin{figure}
\plotone{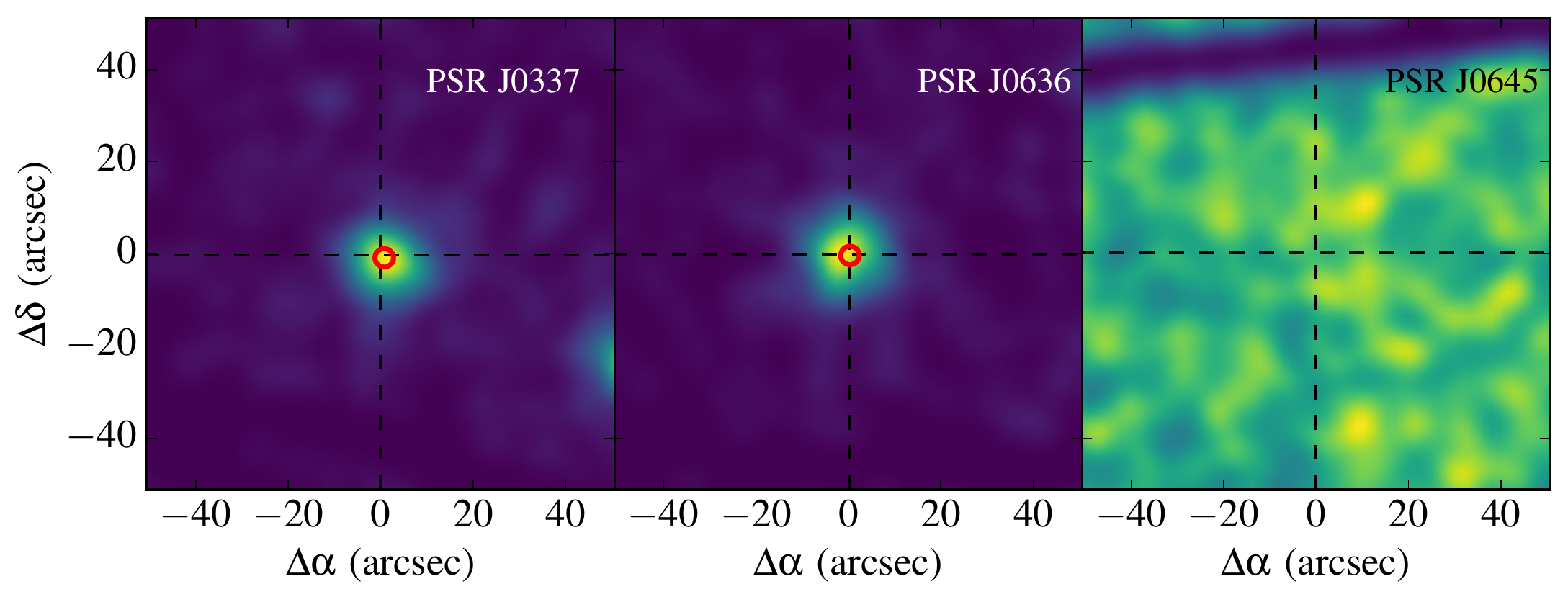}
\caption{X-ray images of PSR\,J0337$+$1715 (left panel), PSR\,J0636$+$5129 (middle panel), and PSR\,J0645$+$5158 (right panel).  Data limited to events with \texttt{PATTERN} $\leq4$ (singles and doubles), with energies between 0.2 and 2.0\,keV.  The black dashed lines indicate the radio positions \citep{rsa+14,slr+14}, and the red circles indicate the $2\arcsec$ uncertainty in the X-ray positions.}
\label{fig:xim}
\end{figure}

\section{Observations and Analysis}
\label{sec:obs}
\subsection{X-ray Data}
\label{ssec:obsx}
Each pulsar system in this analysis was observed with \textit{XMM-Newton} \citep{jla+01} using the European Photon Imaging Camera (EPIC) with pn detector in full frame mode with thin filters (the data from the MOS and RGS detectors had insufficient counts for analysis).  \psrA\ was observed on 2013 August 1 (observation number 0722920101) for 16.2\,ks.  
An X-ray source was detected $1\farcs6$ away from the radio position of \citet{rsa+14}, consistent with the $2\arcsec$ astrometric precision of \textit{XMM}\footnote{See \url{xmm.vilspa.esa.es/docs/documents/CAL-TN-0018.pdf}.}; we show an image of the detection in Figure~\ref{fig:xim}.  We measured 
$164\pm13$ background-subtracted counts between 0.2 and 2.0\,keV, as determined using \texttt{calc\_data\_sum} in \texttt{Sherpa} \citep{fds01,dns+07} and uncertainty given by a Poisson distribution\footnote{Note that our count-rate for \psrA\ is below the 2-sigma upper limit from \citet{pb15}, who analyzed the same data set.  Nonetheless, we are confident in our detection (Fig.~\ref{fig:xim}), and do not know the reason for the discrepancy.}.  
The chance coincidence probability, given the number of sources in the field with similar or higher count rates, is approximately $8\times10^{-5}$.  
\psrB\ was observed on 2013 October 13 (observation number 0722920201) for 15.0\,ks, 
and we found an X-ray source within $0\farcs3$ of the radio position of \citet{slr+14}; see Figure~\ref{fig:xim}. The chance coincidence probability for \psrB\ is also approximately $8\times10^{-5}$.  
We measured $170\pm13$\,counts between 0.2 and 2.0\,keV.  Finally, \psrC\ was observed on 2014 March 29 (observation number 0722920301) for 34.9\,ks, but removing a flare from the data reduced the effective observation length to 23\,ks. 
No source was found by the \textit{XMM} pipeline near the radio position of \citet[][see Fig.~\ref{fig:xim}]{slr+14}, and we measured only $18\pm9$ source counts between 0.2 and 2.0\,keV.  
The time resolution of 73.4 ms was too coarse to detect pulsations at the rotational periods of the pulsars (2.73, 2.87 and 8.85\,ms; \citealt{rsa+14,slr+14}), but the observed flux can guide future searches for pulsed X-rays. 
We reprocessed the data using SAS v13.0.1, specifically \texttt{epchain}.  
Using \texttt{HEAsoft} v6.14 and \texttt{CIAO} v4.6, and some custom scripts, we extracted the source counts from within a radius of $25\,\arcsec$, and background counts from an annular region with radii of $50\,\arcsec$ and $125\,\arcsec$, restricted to the same CCD chip with other sources removed.  We limited the data to events with \texttt{PATTERN} $\leq4$ (singles and doubles) but also experimented with using \texttt{PATTERN} $\leq12$ (singles, doubles, and triples).  We found that the change in the results when including triple events was negligible.  Because of the high background rate at low energies and the expected softness of the source spectra, 
we limited our analysis to energies between 0.2 and 2.0\,keV.  We grouped the counts such that each energy bin 
had at least 15 events in it and subtracted the background from the source.  

\begin{figure}
\plotone{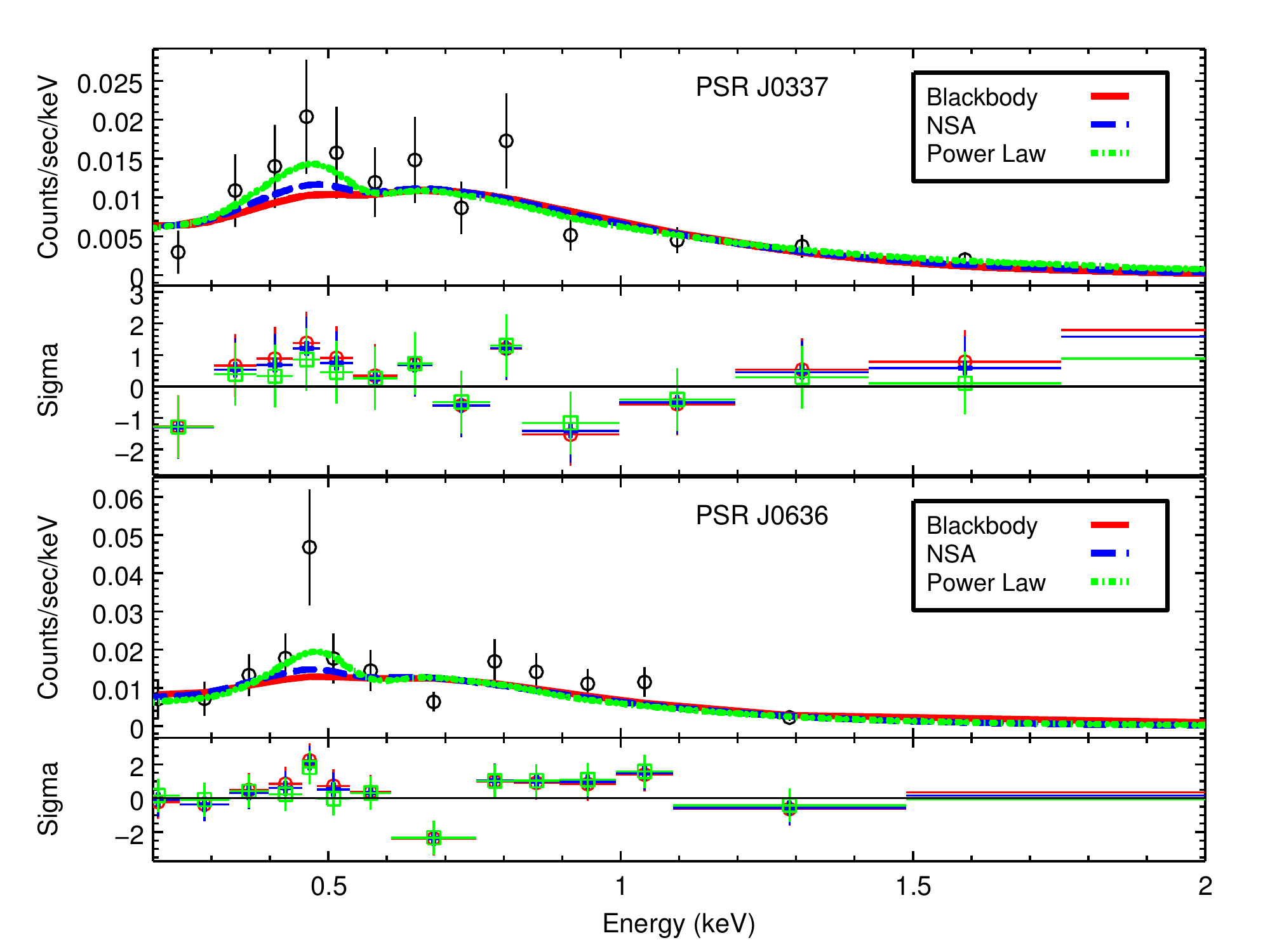}
\caption{X-ray spectra and scaled residuals of PSR\,J0337$+$1715 (top frames) and PSR\,J0636$+$5129 (bottom frames) with free \NH.  The red solid lines are the blackbody model fits; the blue dashed lines are the neutron star atmosphere model fits; and the green dotted lines are the power-law model fits.  See Table~\ref{tab:fit} for the best-fit parameter values and uncertainties.}
\label{fig:mod}
\end{figure}

Using \texttt{Sherpa} \citep{fds01,dns+07}, we fit three models to the data: a power law, a blackbody, and a neutron star atmosphere.  All models also incorporated interstellar absorption using the \texttt{xswabs} model \citep{mmc83}.  For \psrA\ and \psrB, we fit the models with column density, \NH, free and with two fixed values: the first used the pulsar dispersion measure (DM) and the relation between DM and \NH\ found by \citet{hnk13}, while the second used the three-dimensional extinction model of \citet{dcllc03} integrated to the pulsars' distances and converted to \NH\ using \citet{ps95}.  For \psrC, without a significant detection, we only fit with fixed \NH\ (by both methods).  
We fit the models using a $\chi^2$ statistic with the Levenberg-Marquardt minimization method.  We repeated this using the Nelder-Mead Simplex minimization method with the $\chi^2$ statistic, finding the same results.  Due to the low number of counts per bin, we also checked our work using the Cash statistic \citep{cash79}, and the results were consistent with the $\chi^2$ statistic.  The results are shown in Table~\ref{tab:fit}, where the small numbers of counts lead to large uncertainties on the fitted parameters, and the data for \psrA\ and \psrB\ are plotted in Fig.~\ref{fig:mod}.  All models were statistically acceptable.

The power law (PL) model constrains non-thermal emission from the magnetosphere \citep{dkp+12}.  The unabsorbed PL luminosities of \psrA\ and \psrB, as determined by \texttt{calc\_energy\_flux} 
over the range 2.0-10.0\,keV using the PL parameters with fixed \NH\ (using \citealt{dcllc03} and \citealt{ps95}; see Table~\ref{tab:fit}), are $(1.9\pm0.3)\times10^{30}\,{\rm erg\,s}^{-1}$ and $(2.9\pm0.5)\times10^{28}\,{\rm erg\,s}^{-1}$. 
These luminosities correspond to $\approx 5.6\times10^{-5} \dot{E}$ \citep{rsa+14} and $\approx 5.2\times10^{-6} \dot{E}$ \citep{slr+14}.  
When modeling \psrC, we fixed the value for the power-law index $\Gamma$ 
at $2.8$, which is similar to the values found for the other sources, and to values in literature \citep[e.g.,][]{tjs+08,pkgw07}.  From this fit, the 95\% upper limit on the unabsorbed luminosity (over the range 2.0-10.0\,keV) of \psrC\ is $\lsim 3.2\times10^{29}\,{\rm erg\,s}^{-1}$, which corresponds to $\lsim 1.3\times10^{-3} \dot E$ \citep{slr+14}.  

Using the blackbody model (BB) model, for \psrA, we find an inferred radius of $0.2\pm0.1$\,km and temperature of roughly $0.18\pm0.02$\,keV.  The temperature for \psrB\ is also approximately $0.18\pm0.2$\,keV, and the smaller distance, with respect to \psrA, implies a smaller radius, $0.03\pm0.01$\,km.  These results are similar to other MSPs \citep[e.g.,][]{dkp+12} and are consistent with emission from heated polar caps.  For \psrC, we fixed the temperature to 0.2\,keV, and found a 95\% upper limit for the radius of $0.03$\,km, which is comparable to that of \psrB. 

We also fit the data with a neutron star atmosphere (NSA) model (\texttt{xsnsa} in \texttt{sherpa}; \citealt{zps96}).  
For all objects, we set the magnetic field to 0 (appropriate for weakly-magnetized MSPs) and the radius to 15\,km, which is large for a neutron star, but does not significantly affect the resulting emission radii and temperatures.  
The masses were set to $1.438\,M_\odot$ for \psrA\ \citep{rsa+14} and $1.4\,M_\odot$ for \psrB.  We find temperatures of $\approx0.1$\,keV for both pulsars, and emission radii of $1.0\pm0.5$\,km for \psrA\ and $\approx 0.16$\,km for \psrB.  These values are again consistent with heated polar-cap emission from MSPs, where the larger emission radii and lower temperatures, compared to blackbodies, reflect the more realistic hydrogen NSA models. In addition, for \psrC, with the mass set to $1.4\,M_\odot$ and the temperature set to a value similar to those of \psrA\ and \psrB, 0.09\,keV, we find an upper limit for the emission radius of $0.2$\,km.

\subsection{\psrB\ Lightcurve}
\label{ssec:0636}
\begin{figure}
\plotone{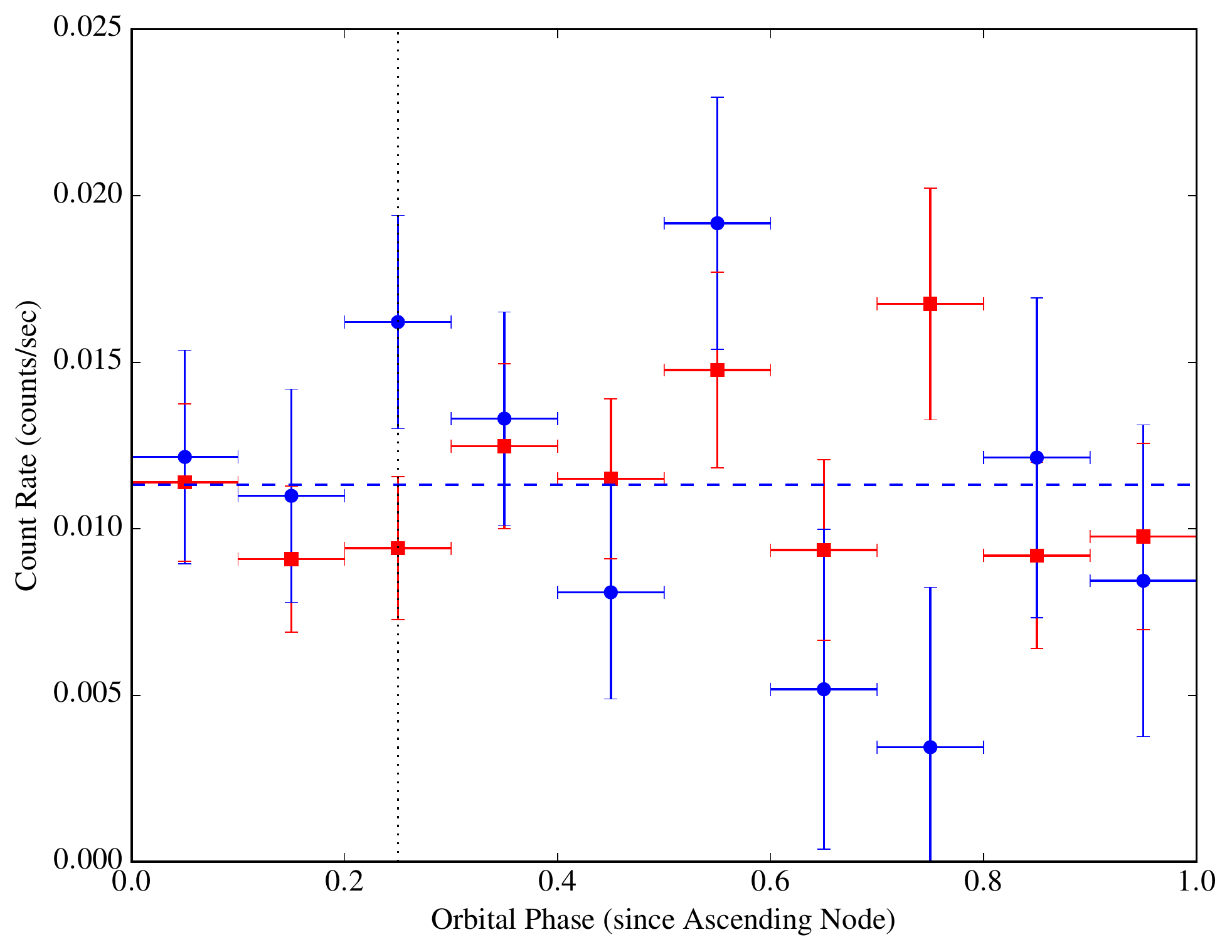}
\caption{PSR J0636$+$5129 lightcurve -- blue circles indicate the background-subtracted source count rate, with mean given by blue dashed line; red squares indicate the count rate from reference source, background-subtracted and scaled to source mean count rate; black dotted line at system conjunction, phase $=0.25$.}
\label{fig:lc}
\end{figure}

Since the X-ray observation of \psrB\ was 15.0\,ks in length and \psrB\ has an orbital period of 5.8\,ks, we checked for significant orbital variation in the data using the ephemeris from \citet{slr+14}. We took the extracted, barycentered event data, subtracted the background, and binned the counts into 10 bins over the orbital period.  We estimated the error in the counts in each bin using the Gehrels approximation of the $\chi^2$ distribution, considering the low numbers of counts in some bins \citep{gehrels86}.  We then scaled the binned counts according to the exposure time for each bin.  We did the same for an unrelated source of similar brightness for comparison.  We found a $\chi^2_{\rm red}$ value of 1.3 ($\chi^2=12$ for 9 degrees-of-freedom) against a constant lightcurve for the pulsar, and 0.8 for the comparison source.  We tried a number of other choices of binning with similar results. Finally, we compared the lightcurves from the two sources and found a $\chi^2_{\rm red}$ value of 1.6. Overall, we find no evidence for orbital variation of \psrB's X-ray flux (see Fig.~\ref{fig:lc}).  We calculate the approximate fractional uncertainty on the sinusoidal amplitude as $\sim\sqrt{2}\sigma_N/N$, where $N$ is the net source counts and $\sigma_N$ is the uncertainty in $N$, and set a $3\sigma$ upper limit of 50\% to any sinusoidal orbital modulation.

\subsection{Optical/UV Data}
\label{ssec:opt}
\begin{figure}
\plotone{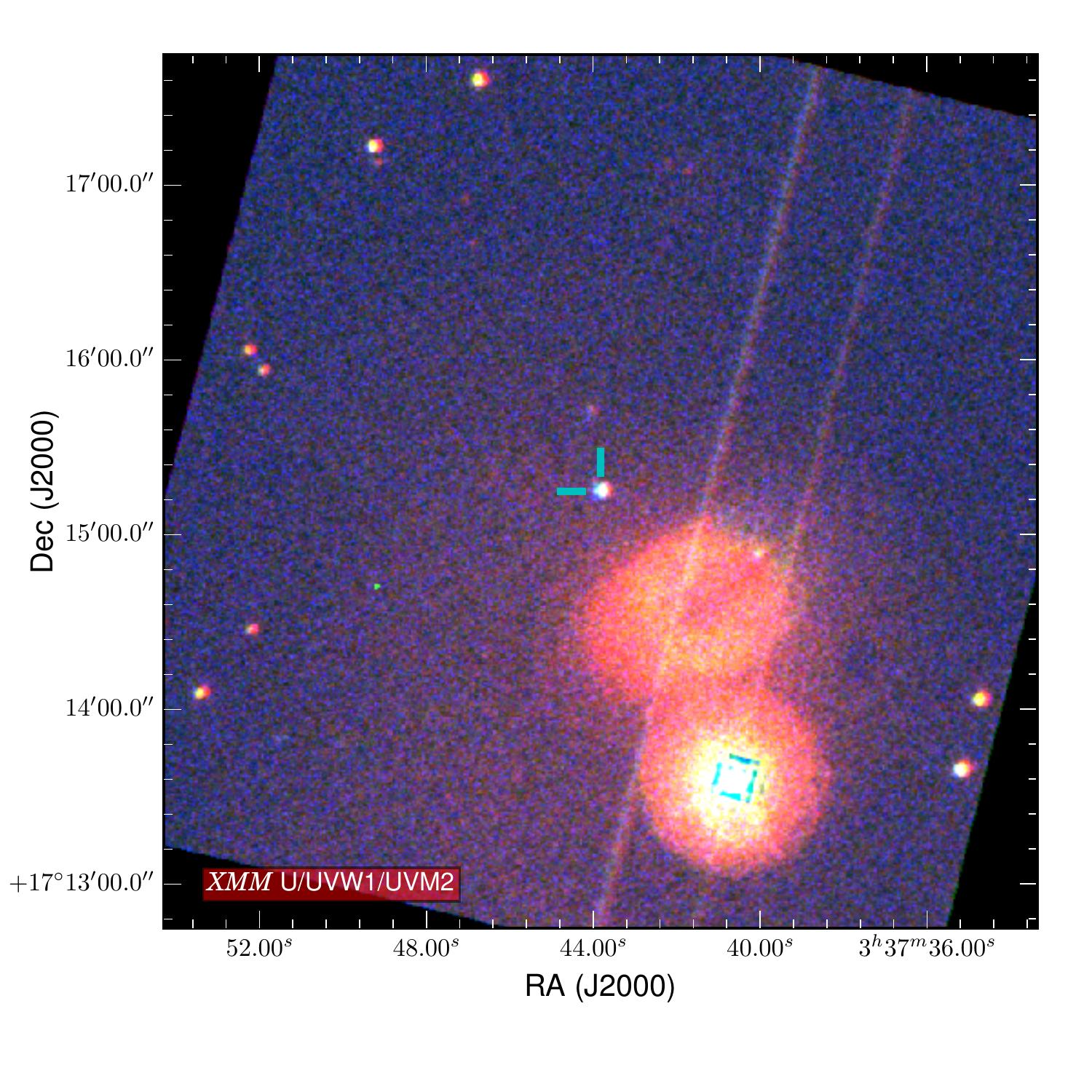}
\caption{3-color composite image of \xmm\ OM data on
  PSR\,J0337$+$1715.  The counterpart is indicated by the tick marks at
  the center.  The image is $5\arcmin$ on each side, with north up and
  east to the left.  The composite is made from $U$ (3440\,\AA),
  $UVW1$ (2910\,\AA), and $UVM2$ (2310\,\AA) observations.   The
  linear streaks are readout trails from the bright star, the diffuse
  circular region is internally-reflected light from that star, and
  the square box indicates a saturated region. }
\label{fig:om1}
\end{figure}

We observed all targets using \textit{XMM-Newton}'s Optical Monitor (OM; \citealt{mbm+01}), but only \psrA\ was detected (see Fig.~\ref{fig:om1}).  
\psrB\ was observed with the $U$ (3440\,\AA) filter for a total exposure time of 13.1\,ks, but was undetected (deep optical/near-infrared searches for \psrB\ will be reported elsewhere).  A bright source nearby may cause some contamination at the radio position, but the effect is minor.  The background noise gives a $3\,\sigma$ limiting magnitude of the system of $21.5$ (AB).  
\psrC\ was observed with the $U$ filter for a total exposure time of 29.6\,ks, but was also undetected, with a $3\,\sigma$ limiting AB magnitude of $21.8$.  
The data on \psrA\ consist of 2 exposures each in the $U$, $UVW1$ (2910\,\AA), and $UVM2$ (2310\,\AA) filters, for total exposure times of 4.7\,ks, 5.88\,ks, and 6.0\,ks, respectively.  We reprocessed the data using SAS 13.5.0 with the latest calibration set, performing point-spread function (PSF) photometry with background regions that accounted for the scattered light halos of nearby stars.  Overall, we find magnitudes relative to Vega (where $m_{V,Vega} = 0.03$) of $m_U=17.59\pm0.04$, $m_{UVW1}=17.14\pm0.04$, and $m_{UVM2}=17.20\pm0.04$.

\begin{figure}
\plotone{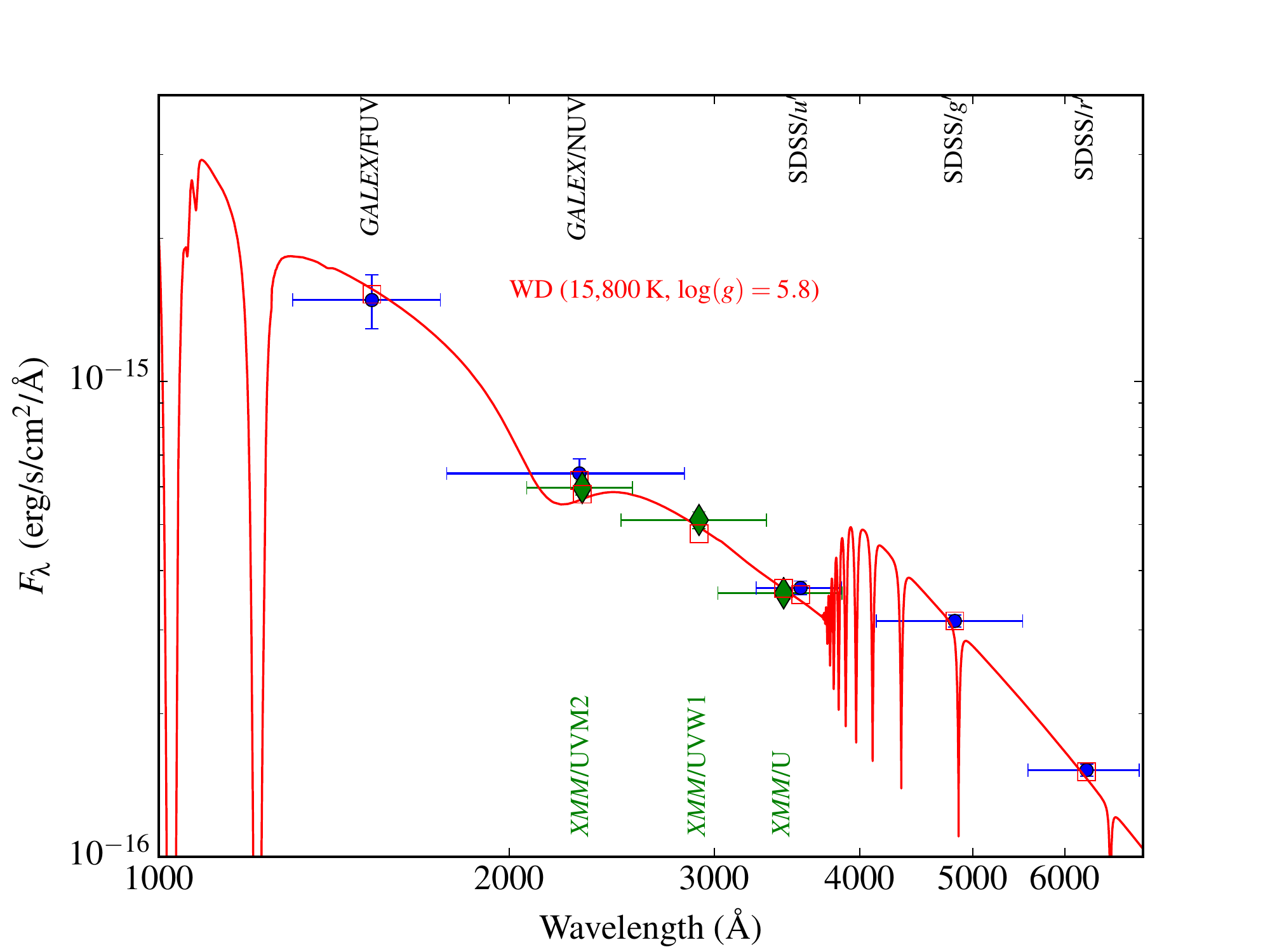}
\caption{Spectral-energy distribution (SED) of the optical counterpart to PSR\,J0337$+$1715.  We show the
optical/ultraviolet portion of the SED presented in \citet{rsa+14} (data shown as blue circles)
with the new \xmm\ OM data shown as the green diamonds.  The
 OM photometric points from each filter are shown, and the red
 squares are the best-fit model atmosphere (including extinction)
 integrated over the filter passband.}
\label{fig:om2}
\end{figure}

\section{Discussion \& Conclusions}
\label{sec:disc}
\begin{figure}
\plotone{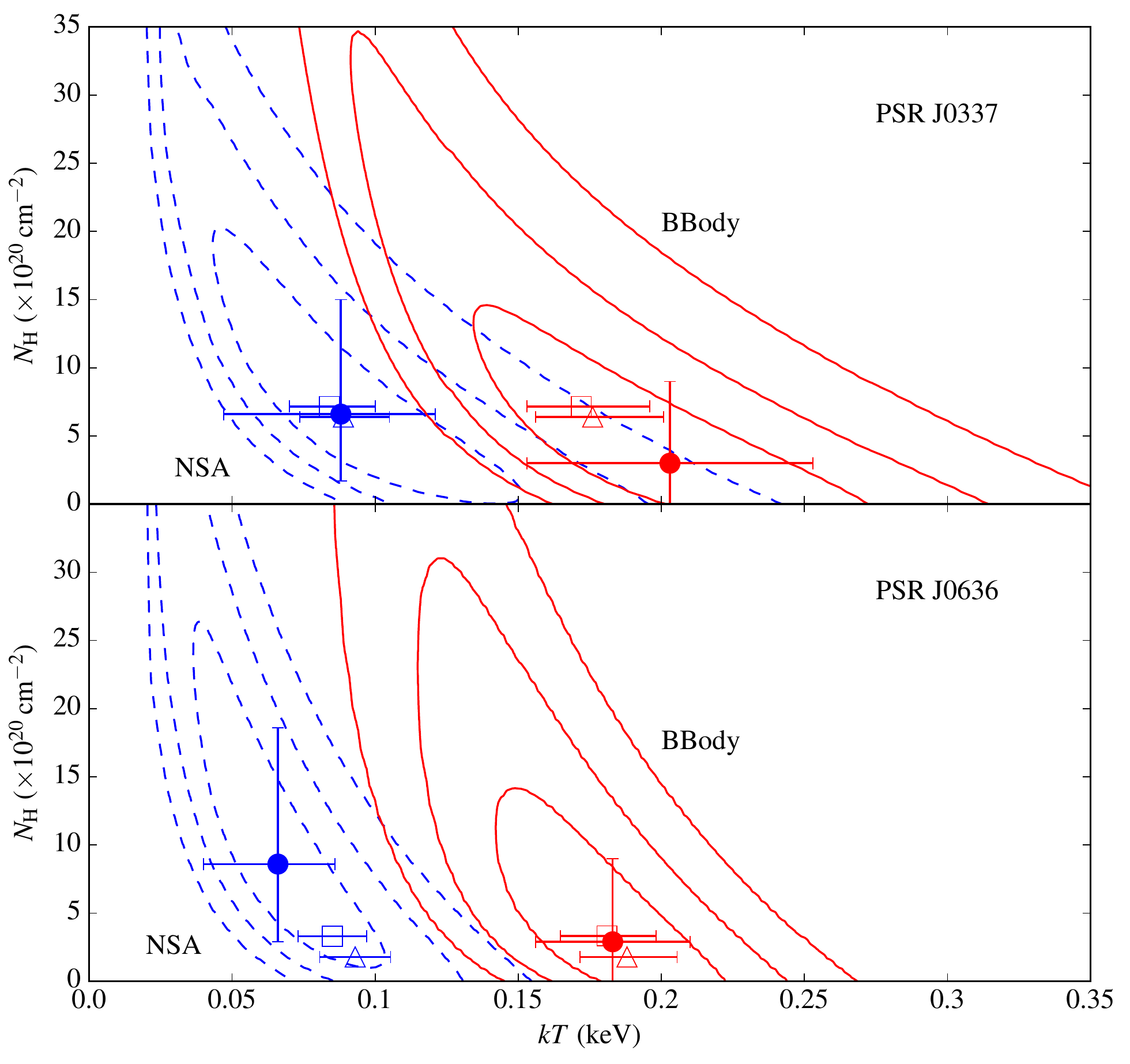}
\caption{Confidence contours for fits to the \xmm\ observations of PSR\,J0337$+$1715 (upper panel) and PSR\,J0636$+$5129 (lower panel).  We show the neutron star atmosphere (NSA; blue dashed lines) and blackbody (red solid lines) models, with contours at 1-, 2- and 3-$\sigma$ values.  Filled circles represent best-fit values for \NH\ and $kT$ with 1-$\sigma$ errorbars, and empty boxes and triangles represent fits with \NH\ fixed using the DM-\NH\ and $A_V$-\NH\ methods, respectively, described in Section~\ref{ssec:obsx} with 1-$\sigma$ errorbars in $kT$.\label{fig:con}} 
\end{figure}

The X-ray spectral models used above are each very simplistic, and do not take into account emission coming from different processes, such as thermal and non-thermal emission from one source.  Other analyses of MSPs with higher signal-to-noise can fit more realistic models to the data \citep[e.g.,][]{zavlin06}.  Moreover, with limited statistics the fit parameters tend to be highly covariant, as shown in Figure~\ref{fig:con}, where higher values for \NH\ lead to more severe constraints on $kT$.

We compare the PL luminosities, described in Section~\ref{ssec:obsx}, with results from other analyses \citep[e.g.,][]{pccm02,lll08,becker09}, noting that the energy ranges differ between analyses.  The general $\dot E$ relation from \citet{pccm02}, for $2-10$\,keV, assumes a high ratio of non-thermal emission:
\begin{equation}\label{eq:pos}
\log(L_X/{\rm erg\,s}^{-1}) = (1.34\pm0.03) \log(\dot E/{\rm erg\,s}^{-1}) - 14.36\pm1.11.
\end{equation}
Using this relation, we expect $\log(L_{X,J0337}/{\rm erg\,s}^{-1}) = 32\pm2$ (the uncertainty in this is derived from that in Equation~\ref{eq:pos}) and $\log(L_{X,J0636}/{\rm erg\,s}^{-1}) = 31\pm2$.  These results, although they are fairly unconstrained, can be compared with the fluxes calculated from the fits to the data to determine the relative thermal and non-thermal emission.  
For \psrA, the measured luminosity and the luminosity from Equation~\ref{eq:pos} are consistent, implying a high ratio of non-thermal to thermal emission.  For \psrB, the luminosity from the data is less than the luminosity from the Possenti relation, which implies the non-thermal emission is less significant.  For \psrC, we use the $\dot E$ from \citet{slr+14} to find $\log(L_{X,J0645}/{\rm erg\,s}^{-1}) = 29\pm1$, which is consistent with the upper limit of $L_{X,J0645} \lsim 3.2\times10^{29}$.  
The relation found by \citet{lll08}, also for $2-10$\,keV, is similarly unconstrained: 
\begin{equation}\label{eq:li}
\log(L_X/{\rm erg\,s}^{-1}) = (0.92\pm0.04) \log(\dot E/{\rm erg\,s}^{-1}) - 0.8\pm1.3.
\end{equation}
From this, we expect $\log(L_{X,J0337}/{\rm erg\,s}^{-1}) = 31\pm2$, $\log(L_{X,J0636}/{\rm erg\,s}^{-1}) = 30\pm2$, and $\log(L_{X,J0645}/{\rm erg\,s}^{-1}) = 29\pm2$.

\begin{figure}
\plotone{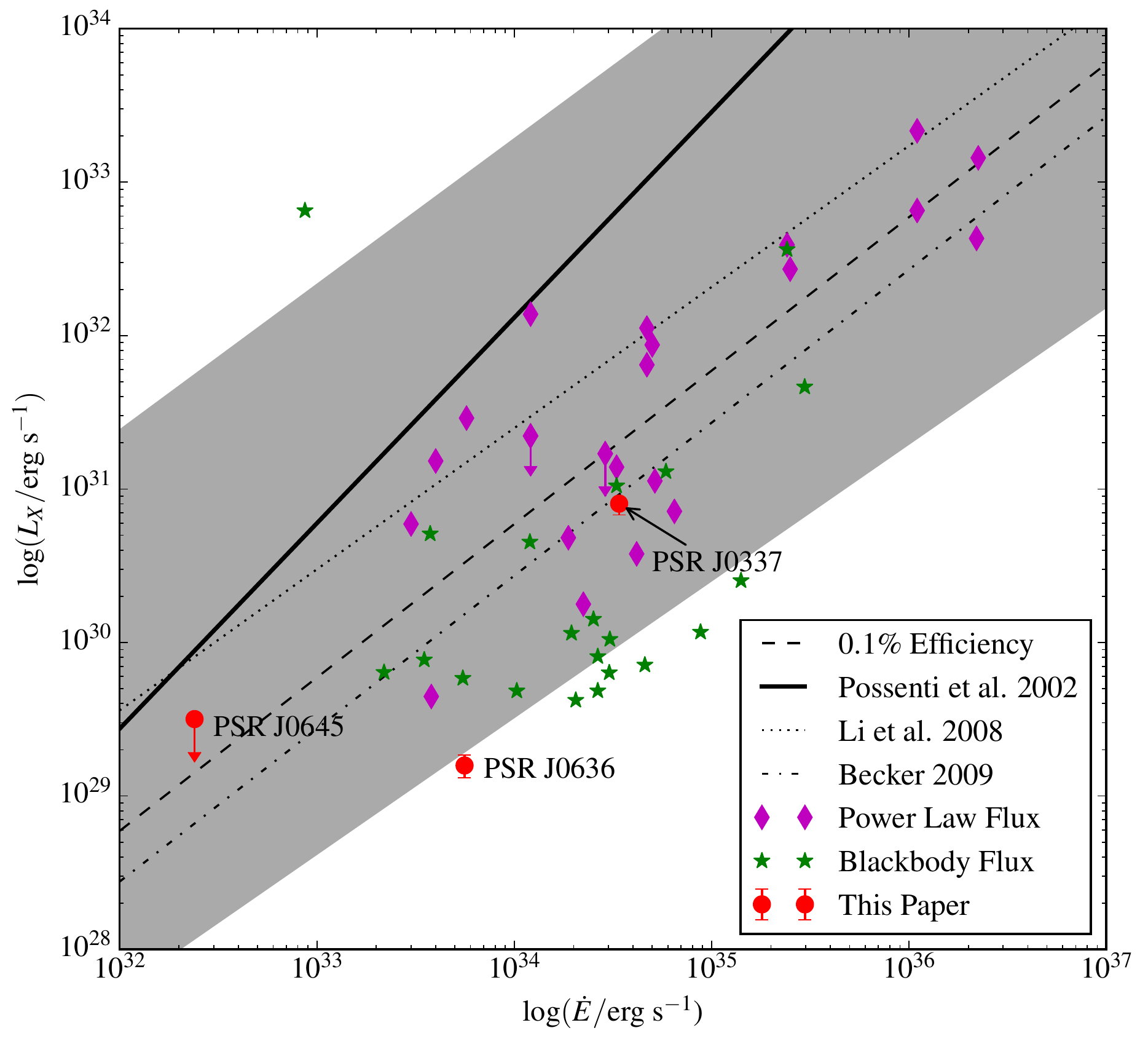}
\caption{X-ray luminosity over $0.2-8.0$\,keV vs. $\dot E$.  Data from \citet{grml+14}, and from this paper (converted to the energy range of $0.2-8.0$\,keV using WebPIMMS).  The solid line is the general relation from Possenti\,et\,al.\,(2002; Equation~\ref{eq:pos}), the dotted line is the relation from Li\,et\,al.\,(2008; Equation~\ref{eq:li}) with the $1\sigma$ uncertainty interval given by the grey region, the dot-dashed line is the relation from \citet{becker09}, and the dashed line represents $0.1\%$ efficiency at $0.1-2.4$\,keV \citep{bt97}.  All relations have been converted to the $0.2-8.0$\,keV energy range.}
\label{fig:gen}
\end{figure}

These results seem consistent with the wider population of MSPs \citep[e.g.,][]{dkp+12}. In Fig.~\ref{fig:gen}, we compare our results with those of other analyses of X-rays from pulsars, compiled by \citet{grml+14}.  The \citet{pccm02} relation, which was formally determined for the $2-10$\,keV range but scaled to the $0.2-8$\,keV range using WebPIMMS\footnote{\url{http://heasarc.gsfc.nasa.gov/cgi-bin/Tools/w3pimms/w3pimms.pl}}, does not fit the data as well as the \citet{lll08} relation (also scaled from $2-10$\,keV to $0.2-8$\,keV), or the simpler $L_X = 10^{-3}\times\dot{E}$ relation suggested by \citet{bt97} for the $0.1-2.4$\,keV range (which is in turn very similar to the updated relation from \citealt{becker09}).  However, \psrA\ and \psrB\ are targets of intensive multi-wavelength campaigns.  We have a precise neutron star mass for \psrA\ (independent of general relativity
; \citealt{rsa+14}), and with the parallax distance from the VLBA, these results can be extrapolated to the wider population.  \psrC\ already has a parallax distance from timing observations, but the neutron star mass is unknown.  
With precise measurements of the luminosities of these and other pulsars, using the more accurate pulsar distance measurements and models that were made since the publication of \citet{pccm02}, a better analysis of the relation between the spin-down luminosity and X-ray luminosity of pulsars can be done \citep[e.g.,][]{pb15}. 

In order to better constrain the emission mechanisms, 
we want to measure the shape of pulsations, which requires much higher time resolution and better sensitivity than achieved in these observations. With the upcoming \textit{NICER} mission, it will be possible to further constrain the masses and radii of \psrA\ and \psrB\ using the known parallax distances \citep{gao12}.  
Based on current data and assuming a pulsed fraction of 25\% \citep[e.g.,][]{bogdanov13}, we estimate $\approx 100\,$ks in order to see pulsations from \psrA\ with either XMM-Newton or NICER at $\approx 5\,\sigma$.  To accomplish \textit{NICER}'s primary goal of constraining the equation-of-state, the radius must be measured to $\approx 3\,\sigma$, which would require a considerably long observation of $\approx70$\,days.

In the optical and ultra-violet, our results on \psrA\ are fully consistent (within 1\,$\sigma$) with previous
photometry and modeling by \citet{kvkk+14} and \citet{rsa+14}.
Specifically, our $U$-band observation agrees with the Sloan Digital
Sky Survey (SDSS) $u^\prime$ data-point, and our $UVM2$-band
observation agrees with the \textit{GALEX}/NUV data-point.  On the
other hand, the $UVW1$ observation has no prior direct confirmation
but fully supports the model of a white dwarf hydrogen atmosphere with
effective temperature 15,800\,K, surface gravity $\log(g)=5.82$,
extinction $A_V=0.45\,$mag, and normalization of $0.091\,R_\odot$ at a
distance of 1300\,pc.   In Figure~\ref{fig:om2}, we show our new
photometry compared to the model atmosphere integrated over the
appropriate filter transmission curves\footnote{Obtained from \url{ftp://xmm.esac.esa.int/pub/ccf/constituents/extras/responses/OM}.}.

For \psrB, we do not detect any emission with the OM, but this is consistent with expectations.  The absence of a companion in the Two Micron All-Sky Survey \citep{scs+06} or Sloan Digital Sky Survey \citep{ewa+11} limited the effective temperature of the companion to $<3000\,$K, based on the  distance used above and a Roche-lobe filling radius of about $0.1\,R_\odot$.  This implies AB magnitude $m_U>28$ (without accounting for extinction, which could make it even fainter), consistent with our limit (also see \citealt{bbb+11} for deeper searches of a similar object).  Dedicated observations focusing on the near-infrared will likely be required to find the companion.  For \psrC, there is no companion and we are limited to searching for just the MSP itself.  Similar searches have been done; e.g., \citet{mb04} use the VLT to search for PSR\,J2124$-$3358, and find a limiting magnitude of $U>26$. 
Our upper limit on the X-ray blackbody would imply AB magnitude $m_U>39$.

\acknowledgements
We thank the anonymous referee for their useful comments. This work is based on observations obtained with \textit{XMM-Newton}, an ESA science mission with instruments and contributions directly funded by ESA Member States and NASA. R.S. and D.L.K. were partially supported by NASA through grant NNX12AO72G. JWTH acknowledges funding from an NWO Vidi fellowship and from the European Research Council under the European Union's Seventh Framework Programme (FP/2007-2013) / ERC Grant Agreement nr. 337062 (``DRAGNET"). MAM is supported by NSF award number \#1211701.  IHS acknowledges support from an NSERC Discovery Grant and from the Canadian Institute for Advanced Research. Apart from the XMMSAS data reduction pipelines provided by \textit{XMM-Newton}, this research has made use of software provided by the Chandra X-ray Center (CXC) in the application packages CIAO and Sherpa.

{\it Facility:} \facility{XMM (EPIC-pn, OM)}

\begin{deluxetable*}{l c c c c c}
\tablewidth{0pt}
\tablecolumns{5}
\tablecaption{X-ray Fits to Sources\label{tab:fit}}
\tablehead{
\colhead{Model} & \colhead{\NH\tablenotemark{a}} & \colhead{$\Gamma$/$kT$\tablenotemark{b}} 
 & \colhead{$A$\tablenotemark{c}/$R$\tablenotemark{d}} & \colhead{$\chi^2$/DOF} & \colhead{Flux\tablenotemark{e}} \\
  & \colhead{($\times10^{20}\,{\rm cm}^{-2}$)} & \colhead{(keV)} & \colhead{}  & \colhead{} & \colhead{($\times 10^{-13}\,{\rm erg\,cm}^{-2}\,{\rm s}^{-1}$)}
  }
\startdata
\sidehead{PSR J0337+1715}
Powerlaw & $18 ^{+11} _{-7}$ & $3.6 ^{+1.1} _{-0.8}$ & $10 ^{+4} _{-2}$ & $7.72/11$ & $1.2^{+0.6}_{-0.4}$  \\
 & $6.4\tablenotemark{f}$ & $2.4 \pm0.2$ & $6.7 \pm0.8$ & $11.1/12$ & $0.31\pm0.05$  \\
 & $7.2\tablenotemark{g}$ & $2.5 \pm0.2$ & $6.9 \pm0.9$ & $10.5/12$ & $0.34\pm0.05$  \\
Blackbody & $3 ^{+6} _{-3}$ & $0.20 \pm0.05$ & $0.13 ^{+0.21} _{-0.04}$ & $14.8/11$ & \dots \\
 & $6.4\tablenotemark{f}$ & $0.18 \pm0.02$ & $0.19 ^{+0.07} _{-0.05}$ & $15.2/12$ & \dots \\
 & $7.2\tablenotemark{g}$ & $0.17 \pm0.02$ & $0.21 ^{+0.08} _{-0.05}$ & $15.3/12$ & \dots \\
NS Atmosphere\tablenotemark{h} & $7 ^{+8} _{-5}$ & $0.09 ^{+0.03} _{-0.04}$ & $1.0 ^{+7.3} _{-0.4}$ & $12.1/11$ & \dots \\
 & $6.4\tablenotemark{f}$ & $0.09 \pm0.02$ & $1.0 ^{+0.6} _{-0.3}$ & $12.1/12$ & \dots \\
 & $7.2\tablenotemark{g}$ & $0.08 \pm0.02$ & $1.1 ^{+0.7} _{-0.3}$ & $12.1/12$ & \dots \\
\sidehead{PSR J0636+5129}
Powerlaw & $30 ^{+60} _{-20}$ & $5 ^{+5} _{-1}$ & $13 ^{+67} _{-5}$ & 15.1/11 & $15 ^{+2e5} _{-7}$ \\
 & $3.3\tablenotemark{f}$ & $2.6 \pm0.2$ & $4.7 \pm0.7$ & $27.0/12$ & $0.25\pm0.04$  \\
 & $1.8\tablenotemark{g}$ & $2.4 \pm0.2$ & $4.3 \pm0.6$ & $30.6/12$ & $0.2\pm0.03$  \\
Blackbody & $3 ^{+6} _{-3}$ & $0.18 \pm0.03$ & $0.03 ^{+0.02} _{-0.01}$ & 17.5/11 & \dots \\
 & $3.3\tablenotemark{f}$ & $0.18\pm0.02$ & $0.028 ^{+0.008} _{-0.005}$ & $17.5/12$ & \dots \\
 & $1.8\tablenotemark{g}$ & $0.19\pm0.02$ & $0.025 ^{+0.007} _{-0.005}$ & $17.5/12$ & \dots \\
NS Atmosphere\tablenotemark{h} & $9 ^{+10} _{-6}$ & $0.07 ^{+0.02} _{-0.03}$ & $0.3 ^{+2.1} _{-0.2}$ & $17.6/11$ & \dots \\
 & $3.3\tablenotemark{f}$ & $0.08 \pm0.01$ & $0.16 ^{+0.10} _{-0.05}$ & $17.5/12$ & \dots \\
 & $1.8\tablenotemark{g}$ & $0.09 \pm0.01$ & $0.12 ^{+0.06} _{-0.04}$ & $18.3/12$ & \dots \\
\sidehead{PSR J0645+5158\tablenotemark{i}}
Powerlaw & 5.5\tablenotemark{f} & 2.8 & $<0.9$ & 3.87/5 & $<0.055$ \\
 & 3.6\tablenotemark{g} & 2.8 & $<0.8$ & 3.88/5 & $<0.026$  \\
Blackbody & 5.5\tablenotemark{f} & 0.20 & $<0.03$ & 4.06/5 & \dots \\
 & 3.6\tablenotemark{g} & 0.20 & $<0.03$ & 4.06/5 & \dots \\
NS Atmosphere\tablenotemark{h} & 5.5\tablenotemark{f} & 0.086 & $<0.2$ & 4.03/5 & \dots \\
 & 3.6\tablenotemark{g} & 0.086 & $<0.2$ & 4.03/5 & \dots
\enddata
\tablenotetext{a}{$ $Listed uncertainties on all parameters are 1-$\sigma$ bounds from \texttt{sherpa}'s \texttt{proj} command, or are derived from those bounds where relevant.}
\tablenotetext{b}{Temperature at infinity for NSA model}
\tablenotetext{c}{$A$ is amplitude $\times10^{-6}$ keV$^{-3}$cm$^{-2}$s$^{-1}$.}
\tablenotetext{d}{$R$ is the emission radius in km for distances of $1300\pm80$\,pc (J0337; \citealt{kvkk+14}), $210^{+30}_{-20}$\,pc (J0636; \citealt{slr+14}), and $650^{+200}_{-130}$\,pc (J0645; \citealt{slr+14}) from the BB amplitude or from scaling the radius and distance from the NSA normalization.}
\tablenotetext{e}{Unabsorbed Flux between 0.2 and 2.0\,keV}
\tablenotetext{f}{\NH\ fixed using DM from \citet{rsa+14} and \citet{slr+14} and relation between DM and \NH\ found by \citet{hnk13}}
\tablenotetext{g}{\NH\ values fixed using relation with $A_V$, found using DM values from \citet{rsa+14} and \citet{slr+14} and results from \citet{dcllc03} and \citet{ps95}}
\tablenotetext{h}{Change in mass used for NSA models produces change in emission radius and temperature that is insignificant compared to uncertainty in fit parameters.}
\tablenotetext{i}{95\% upper limits on single free parameters}
\end{deluxetable*}

\bibliographystyle{apj}

\begin{thebibliography}{}
 \bibitem[Abdo et al.(2009)]{aaa+09}
   Abdo, A. A., Ackermann, M., Ajello, M., et al. 2009, Science, 325, 848
 \bibitem[Bailes et al.(2011)]{bbb+11}
   Bailes, M., Bates, S. D., Bhalerao, V., et al. 2001, Science, 333, 1717
 \bibitem[Becker(2009)]{becker09} 
   Becker, W.\ 2009, Astrophysics and Space Science Library, 357, 91
 \bibitem[Becker \& Tr\"umper(1997)]{bt97}
   Becker, W., \& Tr\"umper, J. 1997, A\&A, 326, 682
 \bibitem[Bogdanov(2013)]{bogdanov13}
   Bogdanov, S. 2013, ApJ, 762, 96
 \bibitem[Bogdanov et al.(2008)]{bgr08}
   Bogdanov, S., Grindlay, J. E., \& Rybicki, G. B. 2008, ApJ, 689, 407 
 \bibitem[Boyles et al.(2013)]{blr+13}
   Boyles, J., Lynch, R. S., Ransom, S. M., et al. 2013, ApJ, 763, 80 
 \bibitem[Burgay et al.(2006)]{bjd+06}
   Burgay, M., Joshi, B. C., D'Amico, N., et al. 2006, MNRAS, 368, 283 
 \bibitem[Cash(1979)]{cash79}
   Cash, W. 1979, ApJ, 228, 939
 \bibitem[Champion et al.(2008)]{crl+08}
   Champion, D. J., Ransom, S. M., Lazarus, P., et al. 2008, Science, 320, 1309
 \bibitem[Chatterjee et al.(2009)]{cbv+09}
   Chatterjee, S., Brisken, W. F., Vlemmings, W. H. T., et al. 2009, ApJ, 698, 250
 \bibitem[Deloye \& Bildsten(2003)]{db03}
   Deloye, C. J., \& Bildsten, L. 2003, ApJ, 598, 1217
 \bibitem[Demorest et al.(2010)]{dpr+10}
   Demorest, P. B., Pennucci, T., Ransom, S. M., Roberts, M. S. E., \& Hessels, J. W. T. 2010, Nature, 467, 1081
 \bibitem[Doe et al.(2007)]{dns+07}
   Doe, S., Nguyen, D., Stawarz, C., et al. 2007, in Astronomical Society of the Pacific Conference Series, Vol. 376, Astronomical Data Analysis Software and Systems XVI, ed. R. A. Shaw, F. Hill, \& D. J. Bell, 543
 \bibitem[Drimmel et al.(2003)]{dcllc03}
   Drimmel, R., Cabrera-Lavers, A., \& L\'opez-Corredoira, M. 2003, A\&A, 409, 205
 \bibitem[Durant et al.(2012)]{dkp+12}
   Durant, M., Kargaltsev, O., Pavlov, G. G., et al. 2012, ApJ, 746, 6
 \bibitem[Eisenstein et al.(2011)]{ewa+11}
   Eisenstein, D.~J., Weinberg, D.~H., Agol, E., et al.\ 2011, \aj, 142, 72
 \bibitem[Freeman et al.(2001)]{fds01}
   Freeman, P., Doe, S., \& Siemiginowska, A. 2001, in Society of Photo-Optical Instrumentation Engineers (SPIE) Conference Series, Vol. 4477, Astronomical Data Analysis, ed. J.-L. Starck \& F. D. Murtagh, 76Ð87
 \bibitem[Fruchter et al.(1988)]{fst88}
   Fruchter, A. S., Stinebring, D. R., \& Taylor, J. H. 1988, Nature, 333, 237 
 \bibitem[Gaensler et al.(2008)]{gmcm08}
   Gaensler, B. M., Madsen, G. J., Chatterjee, S., \& Mao, S. A. 2008, PASA, 25, 184 
 \bibitem[Gehrels(1986)]{gehrels86}
   Gehrels, N. 1986, ApJ, 303, 336
 \bibitem[Gendreau et al.(2012)]{gao12}
   Gendreau, K. C., Arzoumanian, Z., \& Okajima, T. 2012, in Society of Photo-Optical Instrumentation Engineers (SPIE) Conference Series, Vol. 8443, Society of Photo-Optical Instrumentation Engineers (SPIE) Conference Series
 \bibitem[Gentile et al.(2014)]{grml+14}
   Gentile, P. A., Roberts, M. S. E., McLaughlin, M. A., et al. 2014, ApJ, 783, 69
 \bibitem[He et al.(2013)]{hnk13}
   He, C., Ng, C.-Y., \& Kaspi, V. M. 2013, ApJ, 768, 64
 \bibitem[Jacoby et al.(2009)]{jbo+09}
   Jacoby, B. A., Bailes, M., Ord, S. M., Edwards, R. T., \& Kulkarni, S. R. 2009, ApJ, 699, 2009
 \bibitem[Jansen et al.(2001)]{jla+01}
   Jansen, F., Lumb, D., Altieri, B., et al. 2001, A\&A, 365, L1
 \bibitem[Jenet et al.(2006)]{jhvs+06}
   Jenet, F. A., Hobbs, G. B., van Straten, W., et al. 2006, ApJ, 653, 1571
 \bibitem[Kaplan et al.(2014)]{kvkk+14}
   Kaplan, D. L., van Kerkwijk, M. H., Koester, D., et al. 2014, ApJ, 783, L23
 \bibitem[Kaplan et al.(2012)]{ksr+12}
   Kaplan, D. L., Stovall, K., Ransom, S. M., et al. 2012, ApJ, 753, 174
 \bibitem[Kramer et al.(2006)]{ksm+06}
   Kramer, M., Stairs, I. H., Manchester, R. N., et al. 2006, Science, 314, 97
 \bibitem[Li et al.(2008)]{lll08}
   Li, X.-H., Lu, F.-J., \& Li, Z.\ 2008, \apj, 682, 1166 
 \bibitem[Luan \& Goldreich(2014)]{lg14}
   Luan, J., \& Goldreich, P. 2014, ApJ, 790, 82
 \bibitem[Lynch et al.(2013)]{lbr+13}
   Lynch, R. S., Boyles, J., Ransom, S. M., et al. 2013, ApJ, 763, 81
 \bibitem[Manchester et al.(2005)]{mhth05}
   Manchester, R. N., Hobbs, G. B., Teoh, A., \& Hobbs, M. 2005, VizieR Online Data Catalog, 7245, 0
 \bibitem[Manchester et al.(2001)]{mlc+01}
   Manchester, R. N., Lyne, A. G., Camilo, F., et al. 2001, MNRAS, 328, 17
 \bibitem[Mason et al.(2001)]{mbm+01}
   Mason, K. O., Breeveld, A., Much, R., et al. 2001, A\&A, 365, L36
 \bibitem[Mignani \& Becker(2004)]{mb04}
   Mignani, R.~P., \& Becker, W.\ 2004, Advances in Space Research, 33, 616 
 \bibitem[Morrison \& McCammon(1983)]{mmc83}
   Morrison, R., \& McCammon, D. 1983, ApJS, 270, 119
 \bibitem[Pavlov et al.(2007)]{pkgw07}
   Pavlov, G. G., Kargaltsev, O., Garmire, G. P., \& Wolszczan, A. 2007, ApJ, 664, 1072 
 \bibitem[Possenti et al.(2002)]{pccm02}
   Possenti, A., Cerutti, R., Colpi, M., \& Mereghetti, S. 2002, A\&A, 387, 993
 \bibitem[Predehl \& Schmidt(1995)]{ps95}
   Predehl, P., \& Schmitt, J. H. M. M. 1995, A\&A, 293, 889
 \bibitem[Prinz \& Becker(2015)]{pb15}
   Prinz, T., \& Becker, W.\ 2015, arXiv:1511.07713
 \bibitem[Rafikov(2014)]{rafikov14}
   Rafikov, R. R. 2014, ApJ, 794, 76
 \bibitem[Ransom et al.(2011)]{rrc+11}
   Ransom, S. M., Ray, P. S., Camilo, F., et al. 2011, ApJ, 727, L16
 \bibitem[Ransom et al.(2014)]{rsa+14}
   Ransom, S. M., Stairs, I. H., Archibald, A. M., et al. 2014, Nature, 505, 520
 \bibitem[Roberts(2011)]{roberts11}
   Roberts, M. S. E. 2011, in AIP Conf., Vol. 1357, Radio Pulsars: An Astrophysical Key to Unlock the Secrets of the Universe, ed. M. Burgay, N. DÕAmico, P. Esposito, A. Pellizzoni, \& A. Possenti (Melville, NY: AIP), 127Ð130, arXiv:1103.0819
 \bibitem[Romani et al.(2012)]{rfs+12}
   Romani, R. W., Filippenko, A. V., Silverman, J. M., et al. 2012, ApJ, 760, L36 
 \bibitem[Sabach \& Soker(2015)]{ss15}
   Sabach, E., \& Soker, N. 2015, MNRAS, 450, 1716
 \bibitem[Skrutskie et al.(2006)]{scs+06} 
   Skrutskie, M.~F., Cutri, R.~M., Stiening, R., et al.\ 2006, \aj, 131, 1163 
 \bibitem[Stovall et al.(2014)]{slr+14}
   Stovall, K., Lynch, R. S., Ransom, S. M., et al. 2014, ApJ, 791, 67
 \bibitem[Tauris \& van den Heuvel(2014)]{tvdh14}
   Tauris, T. M., \& van den Heuvel, E. P. J. 2014, ApJ, 781, L13
 \bibitem[Torres et al.(2008)]{tjs+08}
   Torres, M. A. P., Jonker, P. G., Steeghs, D., et al. 2008, ApJ, 672, 1079
 \bibitem[van Haaften et al.(2012)]{vhnvj12}
   van Haaften, L. M., Nelemans, G., Voss, R., \& Jonker, P. G. 2012, A\&A, 541, A22
 \bibitem[Verbiest et al.(2012)]{vwc+12}
   Verbiest, J. P. W., Weisberg, J. M., Chael, A. A., Lee, K. J., \& Lorimer, D. R. 2012, ApJ, 755, 39 
 \bibitem[Zavlin(2006)]{zavlin06}
   Zavlin, V. E. 2006, ApJ, 638, 951
 \bibitem[Zavlin(2007)]{zavlin07}
   --. 2007, Ap\&SS, 308, 297
 \bibitem[Zavlin et al.(1996)]{zps96}
   Zavlin, V. E., Pavlov, G. G., \& Shibanov, Y. A. 1996, A\&A, 315, 141
\end{thebibliography}

\end{document}